\documentclass[aps,prd,reprint,groupedaddress,showpacs,preprintnumbers]{revtex4-1}
\pdfoutput=1
\usepackage{graphicx}
\usepackage{amssymb,amssymb}
\usepackage{lineno}
\usepackage{subfigure}

\usepackage[compact]{titlesec}
\titlespacing{\section}{0pt}{8pt}{4pt}
\titlespacing{\subsection}{0pt}{8pt}{4pt}
\titlespacing{\subsubsection}{0pt}{8pt}{4pt}

\newlength{\figwidth}
\def\integLumi{17.3~fb$^{-1}$}

\def\MjjLowerLimitCTEQs8{1.95} 
\def\MjjLowerLimitCTEQEXPECTEDs8{1.74} 
\def\MjjLowerLimitStatOnlyCTEQs8{1.99} 
\def\MjjLowerLimitStatOnlyCTEQEXPECTEDs8{1.83}

\def\ElevenBinChiLambdaMinus{8.1}
\def\ElevenBinChiLambdaPlus{12.0}
\def\ElevenBinChiLambdaMinusExp{8.9}
\def\ElevenBinChiLambdaPlusExp{14.1}

\def\HighestDijetMass{5.5}

\def\Pythia{{\sc Pythia}}

\def\Geant{{\sc Geant4}}
\def\Jimmy{{\sc Jimmy}}
\def\Herwig{{\sc Herwig}}
\def\Powheg{{\sc Powheg}}

\def\NLOJET{{\sc Nlojet++}}

\def\mjj{\ensuremath m_{jj}}

\def\pT{$p_{\rm T}$}

\usepackage{preprintcover} 

\PreprintCoverPaperTitle{Search for New Phenomena in Dijet Angular Distributions in Proton-Proton Collisions at $\sqrt{s} = 8$~TeV Measured with the ATLAS Detector}

\PreprintIdNumber{CERN-PH-EP-2015-067}  

\PreprintCoverAbstract{A search for new phenomena in LHC proton-proton collisions at a center-of-mass energy of $\sqrt{s}=8$~TeV was performed with the ATLAS detector using an integrated luminosity of \integLumi. The angular distributions are studied in  events with at least two jets; the highest dijet mass observed is $\HighestDijetMass$~TeV. All angular distributions are consistent with the predictions of the Standard Model. In a benchmark model of quark contact interactions, a compositeness scale below \ElevenBinChiLambdaMinus~TeV in a destructive interference scenario and \ElevenBinChiLambdaPlus~TeV in a constructive interference scenario is excluded at 95\% CL; median expected limits are \ElevenBinChiLambdaMinusExp~TeV for the destructive interference scenario and \ElevenBinChiLambdaPlusExp~TeV for the constructive interference scenario.}

\PreprintJournalName{Phys.~Rev.~Lett.}

\begin{document}



\title{Search for New Phenomena in Dijet Angular Distributions in Proton-Proton Collisions at $\sqrt{s} = 8$~TeV Measured with the ATLAS Detector}

\author{ATLAS Collaboration}

\affiliation{}

\date{31 March 2015}



\begin{abstract}
A search for new phenomena in LHC proton-proton collisions at a center-of-mass energy of $\sqrt{s}=8$~TeV was performed with the ATLAS detector using an integrated luminosity of \integLumi. The angular distributions are studied in  events with at least two jets; the highest dijet mass observed is $\HighestDijetMass$~TeV. All angular distributions are consistent with the predictions of the Standard Model. In a benchmark model of quark contact interactions, a compositeness scale below \ElevenBinChiLambdaMinus~TeV in a destructive interference scenario and \ElevenBinChiLambdaPlus~TeV in a constructive interference scenario is excluded at 95\% CL; median expected limits are \ElevenBinChiLambdaMinusExp~TeV for the destructive interference scenario and \ElevenBinChiLambdaPlusExp~TeV for the constructive interference scenario. 
\end{abstract}

\maketitle


The search for an internal structure of fermions and new forces that might govern that structure is a major goal of modern particle physics.  The most powerful probes are scattering experiments with large momentum transfer.  Collisions of protons at the Large Hadron Collider (LHC) resulting in two energetic jets of particles (dijets) provide the largest momentum transfer currently available, and therefore the deepest probe.

The angular distribution of jets relative to the beam axis in events with high dijet invariant mass ($m_{jj}$) provides stringent tests of perturbative quantum chromodynamics (QCD)  as well as theories of new phenomena.  QCD calculations predict that dijet production, dominated by $t$-channel gluon exchange in the kinematic region of proton-proton ($pp$) collisions at the LHC, features steeply falling $m_{jj}$ distributions and angular distributions peaked at $|\cos(\theta^*)|=1$, where $\theta^*$ is the polar scattering angle in the two-parton center-of-mass frame.  New phenomena, such as strong gravity~\cite{RandallMeade,Feng:2004} or new interactions~\cite{Baur:1987ga,Baur:1989kv,Eichten:1984eu,Chiappetta1991} typically predict angular distributions which are more isotropic.

Previous studies of dijet angular distributions, at the CERN SPS~\cite{Arnison1986244,Bagnaia1984283}, the 
FNAL Tevatron~\cite{CDF:2009DijetSearch,DZero:2009DijetAng}, the CERN LHC 
at $\sqrt{s}= 7$~TeV~\cite{ATLAS:2010eza,CMS:2010dijetmass,CMS:2010centrality,CMS:2011DijetMassAngle,Aad:2011aj,ATLAS:2013eza}, 
and by the CMS Collaboration at $\sqrt{s}=8$~TeV~\cite{CMS:CERN-PH-EP-2014-261}, have reported results consistent with the  Standard Model (SM).   This Letter reports on studies of dijet angular distributions in $pp$ collisions at \mbox{$\sqrt{s}=8$ TeV} in data with an integrated luminosity of \integLumi\ collected with the ATLAS detector in 2012.

A detailed description  of the ATLAS detector is published elsewhere~\cite{DetectorPaper}. The detector is instrumented 
over almost the entire solid angle around the $pp$ collision point, with layers of tracking detectors, calorimeters, and muon 
detectors.

The jets  are measured using a calorimeter system composed of different detector types covering different regions 
in $\eta$~\cite{coord}
 and depth. The electromagnetic calorimeter is composed of liquid-argon sampling calorimeters using lead as absorber, and is split into a barrel ($|\eta|<1.475$) and two end caps ($1.375<|\eta|<3.2$). The hadronic calorimeter is divided into a barrel and two extended barrels ($ | \eta | < 1.75$), and two endcaps ($1.5<|\eta|<3.2$). The barrel and extended barrels 
are sampling calorimeters with steel as absorber and scintillator tiles as the active medium, while the hadronic endcaps are liquid-argon calorimeters with copper as absorber. In the very forward regions ($3.1<|\eta|<4.9$) there are liquid argon calorimeters with 
copper and tungsten absorbers.

The data are selected using a trigger that requires a single high-$p_{\mathrm{T}}$~\cite{coord} jet above one of eight thresholds, ranging from 25 GeV to 220 GeV.  Due to the high rate of jets at lower $p_{\mathrm{T}}$, only a fraction of the events from the lower seven thresholds are stored.

Individual jets are reconstructed using the anti-$k_t$ jet clustering algorithm~\cite{antikT,Cacciari:2006} with radius
parameter $R = 0.6$. The inputs to this algorithm are clusters~\cite{Aad:2014bia}
of calorimeter cells with energy depositions significantly above the noise. Jet four-momenta are constructed by the 
vectorial addition of clusters of cells, treating each cluster as a four-momentum with
zero mass.  The jet four-momenta are then corrected to the jet energy scale~\cite{ATLAS-CONF-2011-032} as a function 
of $\eta$\ and \pT\ for various effects, the largest of which are the hadronic shower response, detector material 
distribution, and pileup events~\cite{TheATLAScollaboration:2013pia}. This is done using a calibration scheme based on samples of  simulated events and validated with test-beam~\cite{Adragna:2010zz} and collision 
data \cite{Aad:2014bia} studies.

The rapidity of a jet is defined as \mbox{$y = \frac{1}{2}\mathrm{ln}[(E + p_z)/(E - p_z)]$}, where $E$ is the jet energy 
and $p_z$ is the momentum component along the beam axis~\cite{coord}. The scattering angle between two jets can be expressed using the variable \mbox{$\chi=\mathrm{e}^{|y_1-y_2|}=\mathrm{e}^{2|y^*|}$}, where $y_1$ and $y_2$ are the  rapidities 
of the two  jets, and \mbox{$y^*=\frac{1}{2}(y_1-y_2)$}.  The rapidity boost of the dijet system with 
respect to the center of mass of the colliding protons is calculated as $y_{\mathrm{B}}=\frac{1}{2}(y_1+y_2)$.

For each trigger, the event is required to have a jet with  $p_{\mathrm{T}}$ sufficient to achieve a trigger efficiency greater than 99.5\%. For the lowest (highest) threshold trigger, this corresponds to $p_{\mathrm{T}} > p_{\mathrm{T}}^{\mathrm{min}}= 47\ (333)$~GeV. Events are required to have at least two jets, each with $p_{\mathrm{T}} > 50$~GeV; the dijet system, defined as the two jets with largest $p_{\mathrm{T}}$, is required to have $|y^*| < 1.7$, $|y_{\mathrm{B}}| < 1.1$, and $\mjj > 600$~GeV (where the $\mjj$ requirement avoids the kinematic bias in the angular distributions introduced by the minimum $p_{\mathrm{T}}$ requirement).   

The detector covers the angular range $|y^*|<1.7$, corresponding to $\chi < 30$. This interval is divided into 11 bins, with boundaries at $\chi_n = \mathrm{e}^{(0.3 \times n)}$ for $n=0$ to $10$, approximating the segmentation of the calorimeter 
in $\Delta\eta$. The data are further binned coarsely in $\mjj$ with the expectation that low-$\mjj$ bins are dominated by QCD processes and that signals associated with new physics would be found in higher dijet invariant mass bins.  The bin edges are chosen to optimize the expected sensitivity to the  model of contact interactions. The highest dijet mass observed is $\HighestDijetMass$~TeV.

The SM predictions are estimated using the \Pythia8~\cite{pythia8} v8.160 event generator with the AU2~\cite{AU2tune} underlying-event tune and the CT10~\cite{CT10pdf} parton distribution functions (PDF). The simulated events are propagated 
through a detector simulation~\cite{ATLSIMfull} that uses the \Geant~\cite{Agostinelli:2002hh} simulation package.   Pileup conditions vary as a function of the instantaneous luminosity and are taken into account by overlaying simulated minimum-bias events generated with  \Pythia8  onto the hard-scattering process such that the observed distribution of the average number of interactions per bunch crossing is reproduced.
The same reconstruction and event selection are applied to the simulated events and the data.

The \Pythia8 calculations are primarily to leading order (LO) in QCD with simulation of higher-order contributions included in the shower modeling. Events generated by \Pythia8 are reweighted using a correction factor calculated based on the ratio of the next-to-leading-order (NLO) cross-section calculation from \NLOJET~\cite{NLOJETPP1, NLOJETPP2, NLOJETPP3} v4.1.2 to the LO+shower calculation from  \Pythia8:

\[  K(\chi,\mjj) =  \frac{\sigma_{\mathrm{NLO}}(\chi,\mjj)^{\mathrm{NLOJET++}}}{ \sigma_{\mathrm{LO+shower}}(\chi,\mjj)^{\mathrm{\Pythia8}}}. \] 

\noindent
The $K$-factors decrease with $\chi$ and thus modify the shape of the angular distributions; the impact ranges from a few percent at low $\mjj$ to approximately 15\% for the highest $\mjj$ region. Additional processes accounting for electroweak (EW) effects not included in \Pythia8 (virtual weak boson exchange and Sudakov-type logarithms) are included as EW corrections~\cite{Sudakov}.  The effect is most pronounced at high $\mjj$ and low $\chi$, and the correction factors range from unity at low $\mjj$\ to 0.98 -- 1.12 in the highest $\mjj$\ region. The EW corrections and the NLO $K$-factors are applied as a function of $\chi$ and $\mjj$ to the \Pythia8 prediction.

\begin{figure}[h!]
  \centering 
\includegraphics[width=0.5\textwidth]{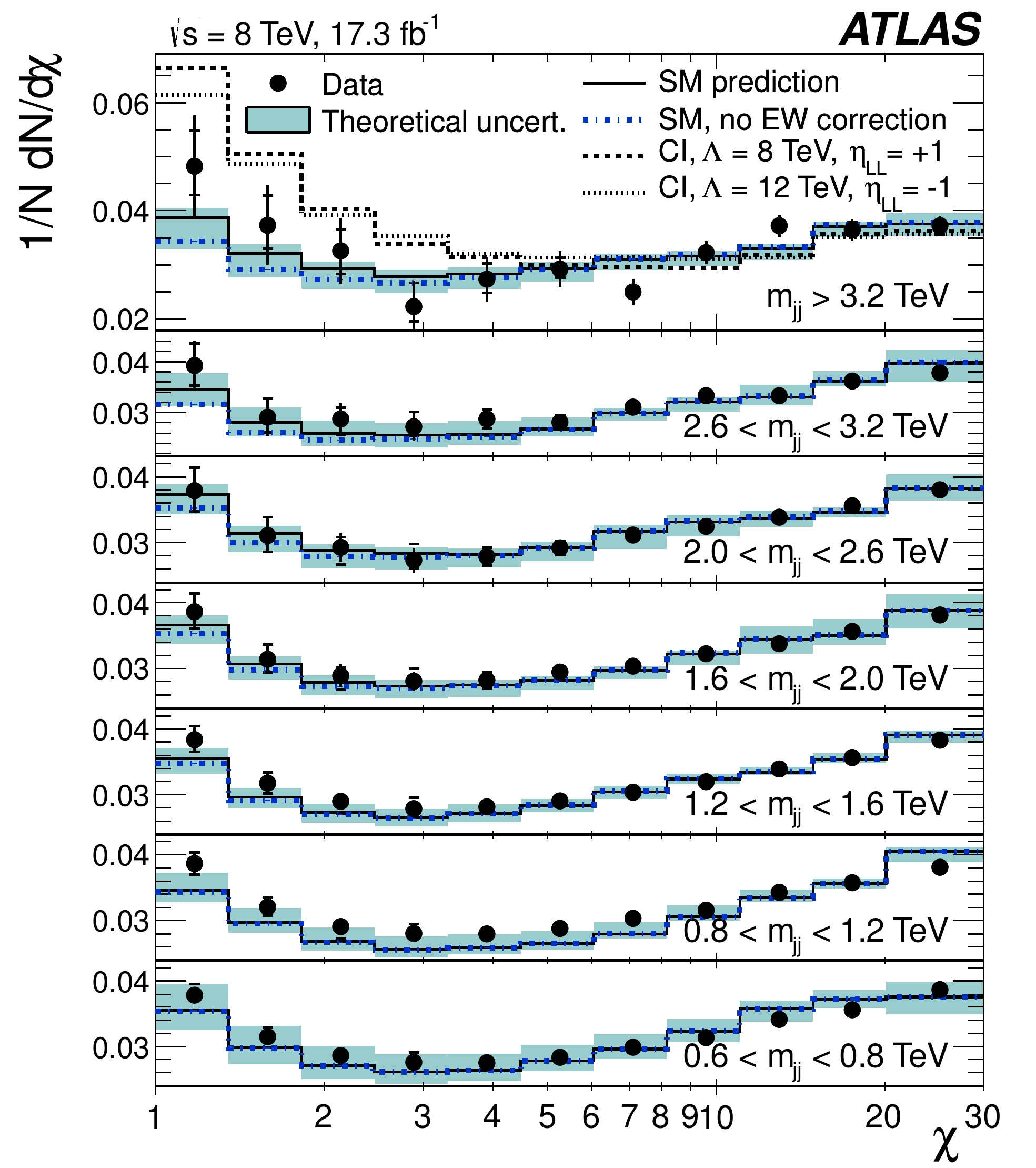}
  \caption{ Distributions of the dijet angular variable $\chi$ for several regions of dijet invariant mass $\mjj$.  Shown are the data (dots),  SM predictions (solid lines) with uncertainties, and predictions for a benchmark theory of new contact interactions with a left-chiral color-singlet coupling with constructive (dotted, $\eta_{\mathrm{LL}}=-1$ and $\Lambda=12$~TeV) or destructive (dashed, $\eta_{\mathrm{LL}}=+1$ and $\Lambda=8$~TeV) interference with the SM processes. Also shown is the impact of the electroweak (EW) corrections described in the text. The error bars on data represent the experimental and statistical uncertainties added in quadrature, with a tick representing experimental uncertainties only. The theoretical uncertainties are displayed as a shaded band around the SM prediction.}
  \label{fig:chimjj}
\end{figure}

Figure~\ref{fig:chimjj} shows the distributions of the data as a function of $\chi$. The distribution in each $\mjj$ region is normalized to unity, as the sensitivity to new phenomena is due to the angular distribution rather than normalization.  The predicted SM distributions are also shown in Fig.~\ref{fig:chimjj} and describe the data well. The EW corrections substantially improve the agreement of the SM prediction with data at high $\mjj$, as can be appreciated from the comparison of the predictions with and without these corrections shown in Fig.~\ref{fig:chimjj}. 

Models of quark compositeness are probed by searching for evidence of new interactions between quarks at a large characteristic energy scale, $\Lambda$.  At energies below this scale, the details of the new interaction and potential mediating particles can be integrated out to form a four-fermion contact interaction model~\cite{Eichten:1984eu,  Chiappetta1991} described by an effective field theory:

\begin{eqnarray} 
L_{qq} =  \frac{2\pi}{\Lambda^2}&[\eta_{\mathrm{LL}}(\bar{q_{\mathrm{L}}}\gamma_\mu
q_{\mathrm{L}})(\bar{q_{\mathrm{L}}}\gamma^\mu q_{\mathrm{L}})\\\nonumber
&  + \eta_{\mathrm{RR}}(\bar{q_{\mathrm{R}}}\gamma_\mu q_{\mathrm{R}})(\bar{q_{\mathrm{R}}}\gamma^\mu q_{\mathrm{R}}) \\\nonumber
&  + 2\eta_{\mathrm{RL}}(\bar{q_{\mathrm{R}}}\gamma_\mu q_{\mathrm{R}})(\bar{q_{\mathrm{L}}}\gamma^\mu q_{\mathrm{L}}) ],
\end{eqnarray}

\noindent
where the quark fields have L and R chiral projections and the
coefficients $\eta_{\mathrm{LL}}$, $\eta_{\mathrm{RR}}$, and $\eta_{\mathrm{RL}}$ turn on and off
various interactions.

In this Letter, a contact interaction (CI) model with a left-chiral color-singlet coupling ($\eta_{\mathrm{LL}}= \pm 1$) is used as a benchmark model, as many other models of new phenomena have similar predictions for the dijet scattering angle $\chi$ at large $\mjj$.   Interference of the signal model with the SM process $q\bar{q}~\rightarrow~q\bar{q}$ is also included.

Event samples were simulated with both QCD and contact interactions, taking interference into account and using the same event generator, underlying-event tune and PDF as for the SM simulations.  Events were generated for both constructive and destructive interference with $\Lambda = 7$~TeV and  $\Lambda = 10$~TeV. The $\Lambda = 7$~TeV sample is then used for extrapolation to other values of $\Lambda$, using the fact that the interference term is proportional to $1/\Lambda^2$ and the pure CI cross section is proportional to $1/\Lambda^4$. This procedure is validated with the $\Lambda = 10$~TeV sample.  As with the QCD prediction, a $K$-factor correction is computed to correct the \Pythia\ LO+shower prediction to an NLO calculation.  Calculations at NLO are provided by  {\sc Cijet}~\cite{CIJET} v1.0.

Uncertainties in the SM and signal predictions include theoretical uncertainties and experimental uncertainties on the measured angular distributions.
Theoretical uncertainties in the SM and signal predictions are due to the choice of PDF, renormalization and factorization scales, choice of event generator as well as statistical uncertainties due to limited simulation sample sizes. The impact of the uncertainty in the PDF is estimated using \NLOJET\ with three different PDF: CT10, MSTW2008~\cite{MSTW2008} and 
NNPDF23~\cite{NNPDF23}. These uncertainties are negligible ($<1\%$), as the choice of PDF largely impacts the total cross section rather than the angular distributions. The uncertainty due to the choice of renormalization and factorization scales were estimated by varying those independently up and down by a factor two in \NLOJET. The resulting uncertainty varies with $\mjj$ and $\chi$, rising to 4\% at the smallest $\chi$ values at high $\mjj$. The uncertainty due to the choice of generator is estimated by comparing the predictions from the NLO generator \Powheg~\cite{POWHEG} v1.0 with those of \Pythia8 with $K$-factors applied. The largest uncertainty due to choice of generator is at the lowest  $\mjj$ values, where it approaches 20\%, while for the highest $\mjj$ values and smallest $\chi$, it ranges from 10\% to 14\%. The uncertainty due to the choice of the showering model is estimated through comparison of \Powheg\ samples showered and hadronized with \Pythia8 v8.175 to \Herwig~\cite{HERWIG} v6.520.2 samples using \Jimmy~\cite{JIMMY1, JIMMY2} v4.31. The largest value of this uncertainty is less than 1\% at the highest $\mjj$ values and smallest values of $\chi$.  Finally the statistical uncertainties on the $K$-factors due to limited simulation sample size are small and set to 1\%. 

The experimental uncertainty is dominated by the $\eta$-dependence of the jet energy scale calibration. This uncertainty 
varies from approximately 15\% at small values of $\chi$ for the highest $\mjj$ values, to a few percent at lower $\mjj$ values and higher 
$\chi$ values. The uncertainty in the beam energy is found to introduce a negligible contribution. The total uncertainty at the lowest $\chi$, highest $\mjj$ amounts to 20\%, decreasing to a few percent at high $\chi$. The total theoretical and experimental uncertainties are shown in  Fig.~\ref{fig:chimjj}.

The $p$-value for the SM hypothesis is (0.25) 0.30  for the (second) highest $\mjj$ bin. In the absence of significant deviations from the SM prediction, upper bounds on CI contributions are calculated using a one-sided profile likelihood ratio and the ${\mathrm{CL}}_S$ technique~\cite{Read:2002hq,Junk:1999kv}, evaluated using the asymptotic approximation~\cite{Cowan:2010js} on events with $\mjj > 3.2$ TeV; the validity of asymptotic approximation was confirmed using toy simulations.  These bounds exclude a compositeness scale below \ElevenBinChiLambdaMinus~TeV in a destructive  interference scenario and below \ElevenBinChiLambdaPlus~TeV in a constructive interference scenario. The median expected limits are \ElevenBinChiLambdaMinusExp~(\ElevenBinChiLambdaPlusExp)~TeV for the destructive (constructive) interference scenario.

In summary, dijet angular distributions have been measured by the ATLAS experiment in \integLumi\ of 8~TeV $pp$ collisions at the LHC.  Over a wide angular range and dijet invariant mass spectrum, the data are well described by QCD predictions at NLO. 
A model of quark compositeness is used as a benchmark for theories of new phenomena that include new forces and mediating particles; such theories predict deviations at small values of $\chi$.  A compositeness scale below \ElevenBinChiLambdaMinus~(\ElevenBinChiLambdaPlus)~TeV in a destructive (constructive) interference scenario is excluded at 95\% confidence level, similar to results from the CMS Collaboration ~\cite{CMS:CERN-PH-EP-2014-261} and representing a significant enhancement in sensitivity relative to the previous limit (at 7.6 TeV for destructive interference) from the ATLAS Collaboration~\cite{ATLAS:2010eza}.


We would like to thank S. Dittmaier and A. Huss for providing us with the electroweak correction factors. We thank CERN for the very successful operation of the LHC, as well as the
support staff from our institutions without whom ATLAS could not be
operated efficiently.

We acknowledge the support of ANPCyT, Argentina; YerPhI, Armenia; ARC,
Australia; BMWFW and FWF, Austria; ANAS, Azerbaijan; SSTC, Belarus; CNPq and FAPESP,
Brazil; NSERC, NRC and CFI, Canada; CERN; CONICYT, Chile; CAS, MOST and NSFC,
China; COLCIENCIAS, Colombia; MSMT CR, MPO CR and VSC CR, Czech Republic;
DNRF, DNSRC and Lundbeck Foundation, Denmark; EPLANET, ERC and NSRF, European Union;
IN2P3-CNRS, CEA-DSM/IRFU, France; GNSF, Georgia; BMBF, DFG, HGF, MPG and AvH
Foundation, Germany; GSRT and NSRF, Greece; RGC, Hong Kong SAR, China; ISF, MINERVA, GIF, I-CORE and Benoziyo Center, Israel; INFN, Italy; MEXT and JSPS, Japan; CNRST, Morocco; FOM and NWO, Netherlands; BRF and RCN, Norway; MNiSW and NCN, Poland; GRICES and FCT, Portugal; MNE/IFA, Romania; MES of Russia and NRC KI, Russian Federation; JINR; MSTD,
Serbia; MSSR, Slovakia; ARRS and MIZ\v{S}, Slovenia; DST/NRF, South Africa;
MINECO, Spain; SRC and Wallenberg Foundation, Sweden; SER, SNSF and Cantons of
Bern and Geneva, Switzerland; NSC, Taiwan; TAEK, Turkey; STFC, the Royal
Society and Leverhulme Trust, United Kingdom; DOE and NSF, United States of
America.

The crucial computing support from all WLCG partners is acknowledged
gratefully, in particular from CERN and the ATLAS Tier-1 facilities at
TRIUMF (Canada), NDGF (Denmark, Norway, Sweden), CC-IN2P3 (France),
KIT/GridKA (Germany), INFN-CNAF (Italy), NL-T1 (Netherlands), PIC (Spain),
ASGC (Taiwan), RAL (UK) and BNL (USA) and in the Tier-2 facilities
worldwide.

\newpage

\bibliography{DijetAngular2012}

\begin{thebibliography}{44}
\bibitem [{coo()}]{coord}%
  \BibitemOpen
  \href@noop {} {}\bibinfo {note} {The ATLAS Collaboration uses a right-handed
  coordinate system with its origin at the nominal interaction point (IP) in
  the center of the detector and the $z$-axis along the beam pipe. The $x$-axis
  points from the IP to the center of the LHC ring, and the $y$-axis points
  upward. Cylindrical coordinates ($r,\phi$) are used in the transverse plane,
  $\phi$ being the azimuthal angle around the beam pipe. The pseudorapidity is
  defined in terms of the polar angle $\theta$ as ${\eta =
  -\ln\left[\tan(\theta/2)\right]}$. The transverse energy and transverse
  momentum are defined by $E_{\rm T} = E\sin{\theta}$ and $p_{\mathrm{T}} =
  |\vec{p}|\sin\theta$, respectively.}\BibitemShut {Stop}%
\end{thebibliography}

\newpage 
\onecolumngrid

\begin{flushleft}
{\Large The ATLAS Collaboration}

\bigskip

G.~Aad$^{\rm 85}$,
B.~Abbott$^{\rm 113}$,
J.~Abdallah$^{\rm 151}$,
O.~Abdinov$^{\rm 11}$,
R.~Aben$^{\rm 107}$,
M.~Abolins$^{\rm 90}$,
O.S.~AbouZeid$^{\rm 158}$,
H.~Abramowicz$^{\rm 153}$,
H.~Abreu$^{\rm 152}$,
R.~Abreu$^{\rm 30}$
% ...
\bigskip
\\
% ...
$^{11}$ Institute of Physics, Azerbaijan Academy of Sciences, Baku, Azerbaijan\\
% ...
$^{30}$ CERN, Geneva, Switzerland\\
% ...
$^{85}$ CPPM, Aix-Marseille Universit{\'e} and CNRS/IN2P3, Marseille, France\\
% ...
$^{90}$ Department of Physics and Astronomy, Michigan State University, East Lansing MI, United States of America\\
% ...
$^{107}$ Nikhef National Institute for Subatomic Physics and University of Amsterdam, Amsterdam, Netherlands\\
% ...
$^{113}$ Homer L. Dodge Department of Physics and Astronomy, University of Oklahoma, Norman OK, United States of America\\
% ...
$^{151}$ Institute of Physics, Academia Sinica, Taipei, Taiwan\\
$^{152}$ Department of Physics, Technion: Israel Institute of Technology, Haifa, Israel\\
$^{153}$ Raymond and Beverly Sackler School of Physics and Astronomy, Tel Aviv University, Tel Aviv, Israel\\
% ...
$^{158}$ Department of Physics, University of Toronto, Toronto ON, Canada\\
% ...
\end{flushleft}


\end{document}